# Icosahedral quasicrystals for visible wavelengths by optical interference holography

Jun Xu[1], Rui Ma[1], Xia Wang[1] & Wing Yim Tam[1]

[1]Department of Physics and Institute of Nano Science and Technology, Hong Kong University of Science and Technology, Clear Water Bay, Kowloon, Hong Kong, China,

Quasicrystals[1,2], realized in metal alloys[3,4], are a class of lattices exhibiting symmetries that fall outside the usual classification for periodic crystals. They do not have translational symmetry and yet the lattice points are well ordered. Furthermore, they exhibit higher rotational symmetry than periodic crystals. Because of the higher symmetry (more spherical), they are more optimal than periodic crystals in achieving complete photonic bandgaps[5,6] in a new class of materials called photonic crystals[7-10] in which the propagation of light in certain frequency ranges is forbidden. The potential of quasicrystals has been demonstrated in two dimensions for the infrared range[11-14] and, recently, in three-dimensional icosahedral quasicrystals fabricated using a stereo lithography method for the microwave range[15]. Here, we report the fabrication and optical characterization of icosahedral quasicrystals using a holographic lithography method[16] for the visible range. The icosahedral pattern, generated using a novel 7-beam optical interference holography, is recorded on photoresists and holographic plates. Electron micrographs of the photoresist samples show clearly the symmetry of the icosahedral quasicrytals in the submicron range, while the holographic plate samples exhibit bandgaps in the angular-dependent transmission spectra in the visible range. Calculations of the bandgaps due to reflection planes inside the icosahedral quasicrystal show good agreement with the experimental results.

Quasicrystalline structures (quasicrystals), discovered in metal alloys in the early eighties, have a higher point group symmetry than ordinary periodic crystals.[3,4] They exhibit long-range aperiodic order and rotational symmetries that fall outside the traditional crystallographic classification schemes.[1,2] It was suggested[5,6] and realized in 2D[11-14] that mesoscale quasicrystals may possess photonic bandgaps (in which electromagnetic wave propagation is forbidden), a character of a new class of materials called photonic crystals[7,8]. Furthermore, because of their higher rotational symmetry, the bandgaps of quasicrystals are more isotropic and thus are more favourable in achieving complete bandgaps than conventional photonic crystals.[7-10] Photonic crystals have been fabricated by techniques such as the self-assembly of colloidal microspheres[17,18] or micro-fabrication[19-20], and, recently, holographic lithography[21-24] and multi-photon direct laser writing[25]. However, it is difficult to fabricate quasicrystals by the self-assembly and micro-fabrication techniques, and is very time consuming to use the direct laser writing technique. Icosahedral quasicrystals exhibiting sizeable bandgaps in the microwave range have been fabricated using a stereo lithography method[15]. However, it is still a challenge to fabricate quasicrystals in the visible range. Recently, it has been demonstrated that holographic lithography can be used to fabricate 2D and quasi-3D quasicrystals[26,27] in photoresists in the submicron range. Furthermore, one of us has shown that it is possible to obtain interference patterns with icosahedral symmetry using a novel 7-beam configuration[28]. Here we report the realization and optical characterization of icosahedral quasicrystals for the visible range using the optical interference holography technique suggested in ref. 28.

Quasicrystals can be classified either as physical 3D projections of higher-dimensional periodic structures or in terms of wave vectors in the reciprocal space corresponding to the diffraction patterns of the quasicrystals[1-2,29]. It is the reciprocal space approach that provides the basis for fabricating quasicrystals using optical interference holography[29]. Figure 1a shows the novel 7-beam interference configuration, with five evenly spaced side beams surrounding two opposite central beams with incidence angle φ, for the icosahedral lattice shown in Fig. 1b. The six lattice basis vectors $\{\vec{a}_i\}$ for the icosahedral quasicrystal are given by Eq. (1) in Table 1. The corresponding face-centered reciprocal basis vectors $\{\vec{q}_i\}$ are given by Eq. (2) in Table 1[29]. One of us has shown that the six reciprocal basis vectors can be generated by seven interfering wave vectors $\{\vec{k}_i\}$ (Eq. (3) in Table 1) as shown in Fig. 1a with φ = 63.4° using Eq. (4) in Table 1[28]. (Note that $\vec{k}_0$ is along the 5-fold axis $\overline{OF}$ in Fig. 1b.) The interference pattern of the seven coherent beams can be written as

$$I(\vec{r}) = \sum_{l,m=0-6} \vec{E}_l e^{-i\vec{k}_l \cdot \vec{r} - i\delta_l} \cdot \vec{E}_m^* e^{i\vec{k}_m \cdot \vec{r} + i\delta_m} , \qquad (5)$$

where $\vec{E}_l$ and $\delta_l$ are the electric field, with amplitude taken as unity, and its phase for wave vector $\vec{k}_l$. We define the polarization of the side beam $\vec{k}_l$ as the angle $\omega_l$ of the electric field $\vec{E}_l$ from the plane of incidence formed by the wave vector $\vec{k}_l$ and the central wave vector $\vec{k}_0$. The polarizations of the central beams $\vec{k}_0$ and $\vec{k}_6$ are taken as the angle from the x-axis on the x-y plane. Figure 1b shows projections of the interference pattern as intensity contour surfaces from Eq. (5) using wave vectors given by Eq. (3) with polarizations $\{\omega_i\} = \{18^o,90^o,90^o,90^o,90^o,90^o,18^o\}$ and equal phases along the 5-fold (F), 3-fold (U), and 2-fold (P) axes, confirming the icosahedral symmetry[28].

Figure 1c shows the experimental setup for the 7-beam interference. Beams, $\{\vec{k}_i\}$ for $i$ = 0 to 5, are obtained by passing an expanded beam from an argon ion laser of wavelength 488 nm through a template with five side holes distributed evenly around a central hole. The six beams with diameter 9 mm, power 15 mW, and polarization adjusted by wave plates mounted on the template, enter a truncated pentagonal pyramid from the base as shown in Fig. 1c. The central beam goes straight up the pyramid, while the side beams are reflected internally at the slanted surfaces of the pyramid. The seventh beam $\vec{k}_6$ is obtained by reflecting $\vec{k}_0$ from a mirror mounted above the sample placed on the truncated pyramid. The side beams, with incidence angle φ, all intersect with the central beams at the sample as shown in Fig. 1c. Using this setup, the beams are more uniform and, more importantly, the phases of the beams are fixed because they are obtained from the same expanded beam. The arrangement in Fig. 1c gives the 5-fold symmetry projection on the sample surface. To obtain other symmetry projections, the sample is sandwiched between a pair of prisms specially made with the appropriate angles as shown in Fig. 1d. Good index matching is essential in the experiment because of the large incidence angle for the side beams.

We used the Shell "SU8" photoresist resin (sensitized to the 488 nm line of the argon ion laser and with refractive index (1.62) the same as the truncated pentagonal pyramid) as the recording medium. The resin was spin-coated on glass substrates with the same refractive index as the SU8 to form ~20 μm thick samples. After exposure of about ~10 s using beam polarizations as in the simulations in Fig. 1b but with no phase adjustment because the icosahedral symmetry is not sensitive to the phases, the sample was developed following the procedures reported earlier[24]. During the developing processes, polymerization occurred only in regions where the exposure dosage exceeded a critical value, while under-exposed regions were washed away, creating a copy of photoresist-air microstructure for the icosahedral quasicrystal. Figures 2b-d and 3b-d show scanning electron microscope (SEM) images of icosahedral samples obtained using two truncated pentagonal prisms for incidence angles φ = 63.4$^o$ and 53.2$^o$, corresponding to the regular (Fig. 2a) and "flattened" icosahedral (Fig. 3a) quasicrystals, respectively. The SEM images show clearly the 5-fold (Figs. 2b and 3b), 3-fold (Figs. 2c and 3c), and 2-fold (Figs. 2d and 3d) symmetries, corresponding respectively to the projections along the F, U, and P directions indicated in Fig. 1b. The upper-left insets are simulated projections of the corresponding symmetries using a lower intensity cutoff than the projections in Fig. 1b. There is a good agreement between experiment and simulations. Note that although the prism angle ($30^o$ chosen from standard optics) used for the incidence angle $\varphi = 53.2^o$ is different from the preferred angles ($31.8^o$ for 3-fold and $26.6^o$ for 2-fold), good 3-fold (Fig. 3c) and 2-fold (Fig. 3d) symmetry projections can still be obtained. The lower-right insets of Figs. 2b and 3b show the 3D nature of the fabricated quasicrystals. Note that Figs. 2c-d, with lattice spacings $\sim$ 200nm, are slightly over exposed, presumably due to the finite resolution of the optical interference system, for attempts with lower exposures were not successful in producing "less-connected" samples. Furthermore, the microstructures of the samples are not sufficiently uniform (due to deformations during developing processes and shrinkage of the photoresist) for optical measurements, even though spectral reflections can be observed at some angles.

In view of the non-uniformity obtained using the SU8 photoresist, we have also used the high-resolution dichromate gelatin (DCG) holographic plate PFG-04 (Slavich, Russia) as the recording medium. However, due to the dielectric mismatch between the gelatin ($\sim$24 μm thick and $n_{DCG} \sim 1.57$ before exposure at 488 nm), the glass substrate (2 mm thick and $n_g = 1.52$), and the truncated pentagonal pyramid ($n_t = 1.62$), we could only obtain good samples with the 5-fold symmetry project on the sample surface using the smaller incidence angle $\varphi = 53.2^o$. (This corresponds to an incidence angle of $\varphi = 55.7^o$ in DCG.) After 15 s exposure, the DCG plate was developed using a thermal hardening procedure refined by us recently to obtained high diffraction effiency[30]. During the developing processes, the well-exposed regions inside the DCG would generate submicron air-voids, confirmed by SEM images[30], while the under-exposed regions would remain intact, creating a 3D microstructure with modulations of refractive index. Since DCG is sensitive to moisture, a cover glass was placed on the top of the developed DCG emulsion and sealed with wax for protection[30]. It was found that the DCG emulsion undergoes an overall 10-20% expansion and 5-10% shrinkage in the z- and xy- directions, respectively, compensating the smaller incidence angle for the regular icosahedral quasicrystal. Despite the relatively low dielectric contrast, the fabricated icosahedral quasicrystal DCG samples (of size ~ 9 mm) show beautiful diffraction patterns projected on the back side of the glass substrate of the DCG sample and sharp reflections under white light as shown in the top schematic diagram of Fig. 4a with the sample facing the incident beam. The right inset of Fig. 4a shows clearly two 5-fold symmetry diffraction orders, inner dark-blue and outer blue-green, corresponding to two different bandgaps at normal incidence, verifying the icosahedral 5-fold symmetry. However, these diffractions, because of the large diffraction angles, are all internally reflected inside the glass substrate and cannot be measured directly by the optical fiber connected to a spectrometer as shown in Fig. 4a, resulting in only one sharp reflection (green line) bandgap at 560 nm and three transmission (red line) bandgaps at 560 nm, 500 nm, and 400 nm for normal incidence as shown in Fig. 4a. The lower-left photo insets of Fig. 4a show the appearance for the normal reflection (green) and transmission (light purple, due to overall high transmittance) using diffused white light, consistent with the optical spectra.

To obtain the angular dependent transmission spectrum, the DCG sample was mounted vertically on a rotation platform as shown in the top inset of Fig. 4a and the incidence angle was scanned every $2.5^o$ in air from $\theta = -65^o$ to $65^o$. This corresponds to $\theta = -37^o$ to $37^o$ in the DCG by using an effective dielectric constant of 1.5 estimated from the volume changes for the DCG after development. Figures 4b and c show transmission bandgaps, as color intensity indicated by the scale bar on the right, versus the incidence angle in the visible range, with the 5-fold symmetry axis along and perpendicular to the axis of rotation, respectively. It can be seen that the bandgaps are symmetric about $\theta = 0^o$ for the 5-fold axis along the rotation axis (Fig. 4b) while they are asymmetric for the perpendicular case (Fig. 4c). Furthermore, the bandgaps are separated into four groups, with three groups corresponding to the three normal transmission bandgaps shown in Fig. 4a, and the fourth one arising only at large incidence angles. It turns out that these bandgaps can be explained by the reflections of planes inside the icosahedral quasicrystal. For every reciprocal vector $\vec{q}_n$ given by $\Delta\vec{k}_{i-j} = \vec{k}_i - \vec{k}_j$ for $i,j = 0$ to 6, planes with spacing given by $2\pi / \left|\Delta\vec{k}_{i-j}\right|$ and normal in the direction of $\Delta\vec{k}_{i-j}$ can be identified from the icosahedral structure inside the DCG. For example, for the reciprocal vector $\vec{q}_0 = \Delta\vec{k}_{0-6}$, the planes with spacing equal to $\lambda/2$ ($\lambda$ is the wavelength of the interfering beam inside the DCG) are parallel to the sample surface[30]. These planes produce the bandgap at 560 nm at normal incidence in Fig. 4a taking into account a 15% swelling in the z-direction, in accord with results reported recently using a 2-beam interference[30]. However, due to the differential effects caused by the swelling in the z- and shrinkage in the xy- directions of the DCG on the planes with different orientations, the spacings and directions of these planes are affected differently. Taking this into account, these planes can then be separated into four groups as those observed in the experiment. After all, the directional bandgaps, drawn as colored solid lines in Figs. 4b and 4c, from the parallel planes produced by the corresponding reciprocal vectors can be calculated using the Bragg condition with a 15% swelling and 6% shrinkage in the z- and xy- directions, respectively, and an effective refractive index 1.5 for the exposed DCG. Although there are missing branches in the experiment—one red and one green in Fig. 4b due to small values of the $\vec{E}_i \cdot \vec{E}_j$ terms in Eq. (5) for the experimental parameters—the agreement with the experimental results is far better than qualitative, confirming further the microstructure of the icosahedral quasicrystals fabricated using the 7-beam optical interference.

To conclude, we have fabricated 3D icosahedral quasicrystals in the submicron range using a novel 7-beam optical interference holographic setup. SEM images of samples fabricated in photoresists show clearly the symmetry of icosahedral quasicrystals, in good agreement with simulated projections of different symmetries. Furthermore, despite the low dielectric contrast, angular dependent transmission spectra with distinct bandgaps are obtained for the first time in the visible range using the DCG samples of icosahedral quasicrystals. The observed branches in the transmission spectra can be explained by the reflections of planes inside the quasicrystals created by the reciprocal vectors constructed from the interfering wave vectors. This work opens new directions in achieving complete bandgaps for photonic crystals using quasicrystals.

**Acknowledgements**  Support from Hong Kong RGC grants CA02/03.SC01, HKUST603303, HKUST603405, and HKUST602606 is gratefully acknowledged.  This work was inspired by C. T. Chan and Ping Sheng.  We thank K. Y. Szeto and N. Wang for helpful discussions.

**Author Information**  Correspondence should be addressed to W. Y. Tam (phtam@ust.hk).  X. Wang is now at the Institute of Photonic Information Technology, School of Mathematics and Physics, QingDao University of Science & Technology, China.

## Table 1: Structural parameters for the icosahedral quasicrystal.

| Lattice base vectors $\{\vec{a}_i\}$ | Reciprocal base vectors $\{\vec{q}_i\}$ | Wave vectors of interfering beams $\{\vec{k}_i\}$ | $\{\vec{q}_n = \Delta\vec{k}_{i-j}\}$ |
|---|---|---|---|
| $\begin{bmatrix} \vec{a}_0 = (0,1,\tau) = \overline{OF} \\ \vec{a}_1 = (\tau,0,1) = \overline{OA} \\ \vec{a}_2 = (1,\tau,0) = \overline{OB} \\ \vec{a}_3 = (-1,\tau,0) = \overline{OC} \\ \vec{a}_4 = (-\tau,0,1) = \overline{OD} \\ \vec{a}_5 = (0,-1,\tau) = \overline{OE} \end{bmatrix} \cdots(1)$ $\tau = (1+\sqrt{5})/2$ is the Golden Mean. | $\begin{bmatrix} \vec{q}_0 = 2\vec{a}_0 = (0,2,2\tau) \\ \vec{q}_1 = \vec{a}_0 + \vec{a}_1 = (\tau,1,1+\tau) \\ \vec{q}_2 = \vec{a}_0 + \vec{a}_2 = (1,1+\tau,\tau) \\ \vec{q}_3 = \vec{a}_0 + \vec{a}_3 = (-1,1+\tau,\tau) \\ \vec{q}_4 = \vec{a}_0 + \vec{a}_4 = (-\tau,1,1+\tau) \\ \vec{q}_5 = \vec{a}_0 + \vec{a}_5 = (0,0,2\tau) \end{bmatrix} \cdots(2)$ | $\begin{bmatrix} \vec{k}_0 = (0,1,\tau) \\ \vec{k}_1 = (-\tau,0,-1) \\ \vec{k}_2 = (-1,-\tau,0) \\ \vec{k}_3 = (1,-\tau,0) \\ \vec{k}_4 = (\tau,0,-1) \\ \vec{k}_5 = (0,1,-\tau) \\ \vec{k}_6 = (0,-1,-\tau) = -\vec{k}_0 \end{bmatrix} \cdots(3)$ | $\begin{bmatrix} \vec{q}_0 = \vec{k}_0 - \vec{k}_6 \\ \vec{q}_n = \vec{k}_0 - \vec{k}_n, n=1-5 \end{bmatrix} \cdots(4)$ |

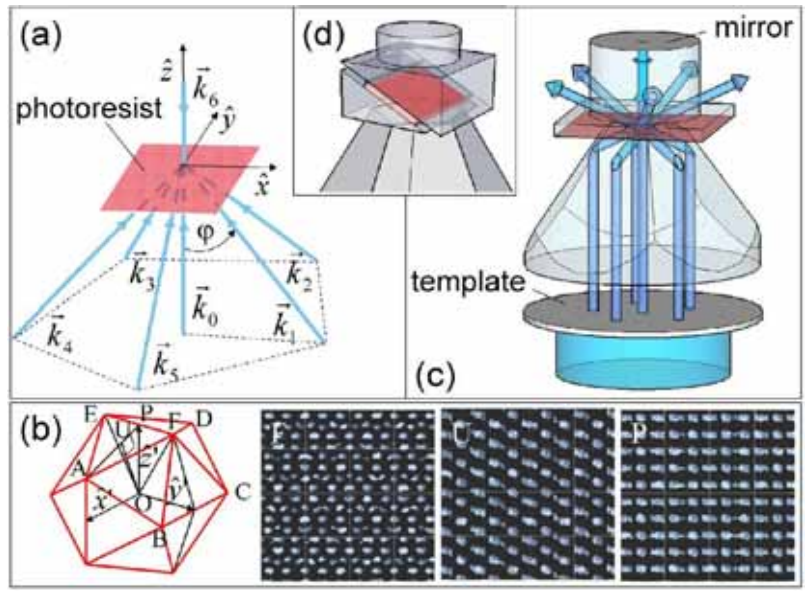

**Figure 1. Beam configuration for icosahedral quasicrystals. a**, 7-beam configuration for the icosahedral quasicrystal. **b**, Icosahedral quasicrystal lattice (red) with $\varphi = 63.4°$ and the simulated 5-fold (F), 3-fold (U), and 2-fold (P) symmetry projections using 70% intensity cutoff. **c**, Actual 7-beam arrangement (for 5-fold symmetry) using a truncated pentagonal prism. **d**, Setup for obtaining 2-fold and 3-fold symmetry projections on the surface of photoresist using a pair of prisms.

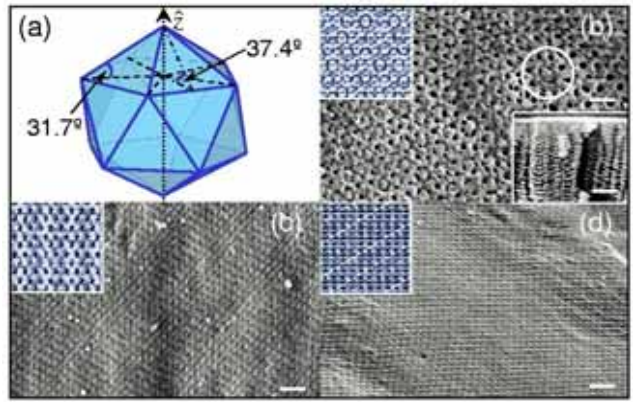

**Figure 2.  Electron micrographs of fabricated icosahedral quasicrystals using incidence angle φ = 63.4° in SU8**.  **a**,  Icosahedral quasicrystal lattice using φ = 63.4°.  **b**, SEM image of the 5-fold symmetry obtained using configuration in Fig. 1b. **c**, SEM image of the 3-fold symmetry obtained using configuration in Fig. 1c with prism angle 37.4°. **d**, SEM image of the 2-fold symmetry obtained using configuration in Fig. 1c with prism angle 31.7°.  The upper-left insets in b-d are simulated projections of the corresponding symmetries using a 40% intensity cutoff. The circle in b shows the 5-fold symmetry.  The lower-right inset in b is the cross-sectional SEM image of the sample. The scale bars are all 1 μm.

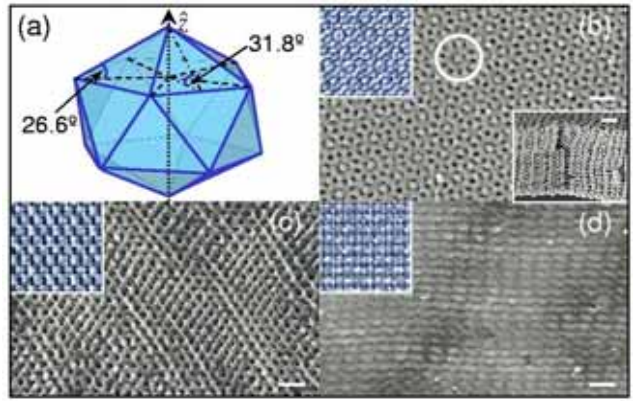

**Figure 3. Electron micrographs of fabricated icosahedral quasicrystals using incidence angle φ = 53.2$^{o}$ in SU8**. **a**, Icosahedral quasicrystal lattice using φ = 53.2$^{o}$. **b**, SEM image of the 5-fold symmetry obtained using configuration in Fig. 1b. **c**, SEM image of the 3-fold symmetry obtained using configuration in Fig. 1c with prism angle 30$^{o}$. **d**, SEM image of the 2-fold symmetry obtained using configuration in Fig. 1c with prism angle 30$^{o}$. The upper-left insets in b-d are simulated projections of the corresponding symmetries using a 40% intensity cutoff. The circle in b shows the 5-fold symmetry. The lower-right inset in b is the cross-sectional SEM image of the sample. The scale bars are all 1 μm.

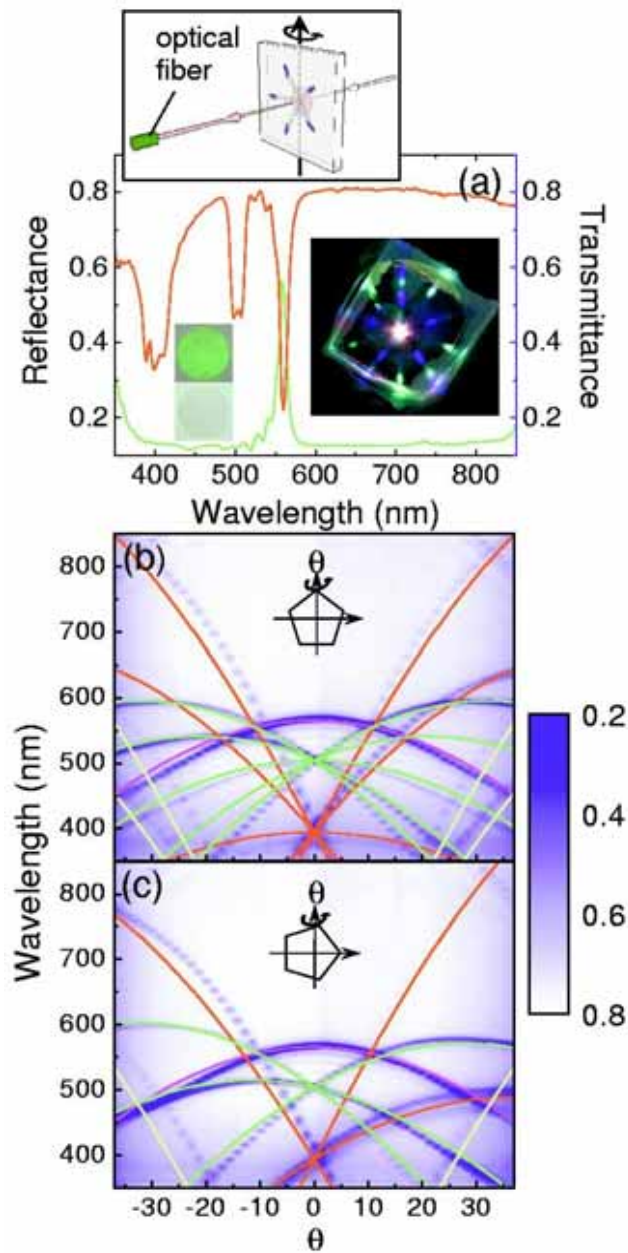

**Figure 4.  Optical measurements of fabricated icosahedral quasicrystals using incidence angle φ = 55.7° in DCG gelatin**.  **a**, Normal reflectance (green) and transmittance (red) of the DCG icosahedral quasicrystal sample using the setup in the top inset.  Right inset is the diffraction pattern of the icosahedral quasicrystal. Lower-left insets are photos of the normal reflection (top green) and transmission (bottom light purple) from diffuse white light.  **b** and **c,** Angular transmission spectra for rotation along and perpendicular to the

5-fold axis, respectively. The scale bar on the right is the transmittance. The color lines are bandgaps of reflection planes inside the icosahedral quasicrystal obtained by the reciprocal vectors $\Delta\vec{k}_{0-6}$ (magenta), $\Delta\vec{k}_{i-6}$ (green), $\Delta\vec{k}_{0-j}$ (red), and $\Delta\vec{k}_{i-j}$ (yellow), for $i,j$ = 1-5.